\documentclass{article}
\usepackage[dvips]{graphicx}
\usepackage{latexsym}
\begin{document}

\title{One-dimensional spinless fermion model with competing interactions beyond
half-filling}
\author{A. K. Zhuravlev and M. I. Katsnelson}
\thanks{Institute of Metal Physics, Ekaterinburg 620219, Russia}
\maketitle

\begin{abstract}
An accurate numerical consideration of 1D spinless fermion model with
next-nearest neighbour (NNN) interactions is carried out for the electron
concentrations 4/7. It is shown that depending on the parameters of the
model it can be either Luttinger liquid or bipolaron liquid. In the former
case competing interactions can result in a smooth behavior of one-electron
distribution function at the Fermi surface with a divergence in the second
derivative with respect to the quasimomentum (and not in the first one, as
usual). In this connection, uncommon photoemission spectra for some 1D
conductors can be explained qualitatively.
\end{abstract}

PACS 71.10.Pm, 71.10.Fd

Frustrations and competing interactions are very important for strongly
correlated electron systems. As it was shown in a seminal work\cite{Fazekas}
they can change drastically a ground state of quantum antiferromagnets
leading to the formation of ``resonating valence bond'' (RVB), or quantum
spin liquid, state (see also \cite{anderson} and references therein). This
state has been found later in a series of compounds with both magnetic (spin
liquid) and charge (pseudospin liquid) degrees of freedom, see, e.g., the
discussion in Refs.\cite{IK,lacroix}.The effects of competing interactions
were investigated recently for one-dimensional (1D) spinless fermion model%
\cite{ZKT,TE,ZK}. Usually this model is used for the description of Verwey
metal-insulator transitions and charge ordering phenomena in Fe$_3$O$_4$, Ti$%
_4$O$_7$ and other d-metal compounds\cite{verwey,verwey1,Kobayashi}. It is
expected that in the half-filled case the growth of the Coulomb interaction
leads to the transition from the metallic state to the insulating
charge-ordered one. It appeared, however, that the competing interactions
can result in the stabilization of metallic phase without charge ordering
for arbitrarily strong Coulomb interaction\cite{ZKT,TE}, this phase being
not usual (for 1D systems) Luttinger liquid\cite{ZK}. Here we investigate
this model beyond half-filling.

We proceed with the Hamiltonian
\begin{eqnarray}
H=-t\sum_{i=1}^L\left( c_i^{\dagger }c_{i+1}+c_{i+1}^{\dagger }c_i\right)
+V\sum_{i=1}^Ln_in_{i+1}+ V^{\prime }\sum_{i=1}^Ln_in_{i+2}
\label{Hsp2}
\end{eqnarray}
where $c_i^{\dagger },c_i$ are Fermi creation and annihilation operators on
site $i,$ $n_i=c_i^{\dagger }c_i$. Further we use periodic boundary
conditions ($L$ is the length of the ring) and put $t=1$ at the presentation
of the computational results.

Earlier we investigated the cases $\rho =1/2$ and $\rho =2/3$ where $\rho
=N/L$ is the electron concentration, $N$ is the number of electrons\cite
{ZKT,ZK}. In that cases the ground state is charge-ordered and insulating
for small enough $t$, except special parameter ratios\ ($V=2V^{\prime }$ for
$\rho =1/2$ and $V=0$ or $V^{\prime }=0$ for $\rho =2/3$) when the ground
state has macroscopically large degeneracy $g=\exp (LS)$, $S$ is the entropy
per cite in the limit $t=0$. Consider now a case $1/2<\rho <2/3$. It is easy
to demonstrate that $S$ is finite at $t=0$ and arbitrary positive $V$ and $%
V^{\prime }$. Therefore one can expect that the ground state is gapless
(conducting) at any small but non-zero $t$. However, the character of the
electron motion turns out to be different for $V>2V^{\prime }$ and $%
V<2V^{\prime }$. In the former case the processes of the motion of a single
electron without the change of the energy of Coulomb repulsion are possible
but in the latter one the electrons can move only as a coupled pairs on
neighbouring sites (bipolarons). For example, the configurations
[...011011001...] and [...011001101...] which are distinguished by the
motion of one bipolaron in right direction have the same interaction energy
(here 1 and 0 are labels for occupied and empty sites). At the same time,
the configurations [...010111001...] or [...011010101...] have the
additional energy equal to $V^{\prime }$ and $2V^{\prime }-V$,
correspondingly. The latter is the value of bipolaron binding energy at $%
t=0,V^{\prime }<V<2V^{\prime }.$ Note that this bipolaron state is similar
to Shubin-Vonsovsky ``maximum polarity'' state\cite{maxpol}. Thus, the cases
$V>2V^{\prime }$ and $V<2V^{\prime }$ are physically different and has to be
considered separately. We investigate both of them numerically for the case $%
\rho =4/7$. Parameters of the ground state, correlation functions, etc. are
calculated by the Lanczos method with the extrapolation to the infinite
length of the chain $L\rightarrow \infty $; the computational methods are
described in Refs.\cite{ZKT,ZK}.

Consider first the region $V>2V^{\prime }$. One can assume that the system
is Luttinger liquid (LL)\cite{haldane} in this case. There are three
branches of low-energy excitations in LL: density fluctuation, current, and
charge excitations with the velocities $v_S,v_J$ and $v_N$, correspondingly.
The first one is connected with the variation of the total energy $E$ of the
system under the variation of the total momentum $P,$ $v_S=\delta E/\delta P$%
; the second one ($v_J$), with the variation of the energy under a shift of
all the particles in momenta space, and the third one, with the variation of
the chemical potential $\mu $ under the change of the total number of
particles $N,$ $v_N=\left( L/\pi \right) \delta \mu /\delta N$. The velocity
$v_J$ is connected with the conducting properties of the system\cite{kohn,SS}%
: $v_J=2\pi D$ where $D$ is the Drude weight (a spectral weight of the
zero-frequency peak in the conductivity). Thus $v_J\neq 0$ is one of the
criterion of the metallic state\cite{kohn}. As it was shown by Haldane\cite
{haldane}, the identity
\begin{equation}
(v_S)^2=v_Jv_N  \label{vvv}
\end{equation}
takes place. Therefore it is possible to introduce the parameter
\begin{equation}
e^{-2\varphi }=v_N/v_S=v_S/v_J  \label{expfi}
\end{equation}
which determines the time and space asymptotics of the correlation functions
of the system. One-electron distribution function $n(k)$ as a function of
the quasimomentum $k$ has a power-low singularity near the Fermi surface, $%
k=k_F$%
\begin{equation}
n(k)\approx n(k_F)-C\textrm{sgn}(k-k_F)|k-k_F|^\alpha .  \label{n(k)}
\end{equation}
with $\alpha =\cosh 2\varphi -1.$ One-particle local spectral density $\rho
\left( E\right) $ has a singularity near the Fermi energy $\mu $ with the
same exponent, $\rho \left( E\right) \propto \left| E-\mu \right| ^\alpha $
which can be measured therefore by photoemission spectroscopy. There are
various experimental data for inorganic materials with charge-density-waves%
\cite{dar1,dar2,hwu} as well as for the organic superconductor (TMTSF)$_2$PF$%
_6$\cite{dar3} which confirms power-law singularity in $\rho \left( E\right)
$ predicted by LL theory. It is interesting that in the latter case the
value $\alpha \simeq 1.25$ has been found. At the same time, in all the
known models with short-range interactions we always have $\alpha <1$ (in
particular, in the 1D Hubbard model one can prove that $\alpha \leq 1/8$\cite
{hubbard}). Our numerical calculations show that $\alpha <1$ is also in the
model under consideration for one-half - or two--third- fillings\cite{ZK}.
Here we demonstrate that $\alpha $ can be larger that 1 for $1/2<\rho <2/3$
which give a qualitative explanation of the experimental data\cite{dar3}.

The computational results are presented in the Table.

\begin{table}[htb]
\caption{ Velocities $v_J,v_N,$ and $v_S$ extrapolated to $L\to\infty$,
the ratio $\chi= v_Jv_N/\left( v_S\right) ^2,$ and the parameter $\alpha $
from Eq.(\ref{n(k)}) for different $V,V^{\prime }$ ($t=1$). For the lower part
of the Table there is no Luttinger liquid state ($\chi, v_N\to\infty$ at
$L\to\infty$ and $\alpha$ is not defined).}
\label{v_J4x7}
\begin{tabular}{|c|c|c|c|c|c|c|}
\hline
$V$ & $V^{\prime}$ & $v_J$ & $v_N$ & $v_S$ & $\chi$ & $\alpha$ \\ \hline
2 & 0   & 1.798  & 4.747  & 2.924 & 0.998 & 0.120  \\
2 & 0.5 & 1.877  & 4.591  & 2.947 & 0.992 & 0.101  \\
2 & 1   & 1.907  & 4.396  & 2.920 & 0.983 & 0.088  \\
2 & 1.5 & 1.889  & 4.328  & 2.892 & 0.978 & 0.088  \\
10 & 0  & 1.074  & 8.059  & 2.939 & 1.002 & 0.572  \\
10 & 5  & 1.668  & 9.314  & 3.806 & 1.065 & 0.408  \\
40 & 11 & 0.886  & 19.582 & 4.197 & 0.985 & 1.456  \\
80 & 22 & 0.816  & 20.848 & 4.132 & 0.997 & 1.624  \\
120 & 33 & 0.793 & 21.301 & 4.108 & 1.001 & 1.686  \\ \hline
10 & 10 & 0.269  &        & 0.757 &       &        \\
10 & 15 & 0.124  &        & 0.379 &       &        \\
10 & 20 & 0.070  &        & 0.248 &       &        \\
10 & 40 & 0.030  &        & 0.103 &       &        \\
10 & 60 & 0.019  &        & 0.065 &       &        \\
20 & 20 & 0.104  &        & 0.377 &       &        \\
40 & 40 & 0.052  &        & 0.182 &       &        \\
80 & 80 & 0.026  &        & 0.089 &       &        \\
100 & 60 & 0.107 &        & 0.367 &       &        \\
\hline
\end{tabular}
\end{table}
The finiteness of $v_J$
and, consequently, of the Drude weight proves the metallic nature of the
ground state. One can see that for $V>$ $2V^{\prime }$ Eq.(\ref{vvv}) is
valid (but not for $V<2V^{\prime }$ where the system is not LL but bipolaron
liquid, see below). The exponent $\alpha $ appears to be maximum along the
line $V^{\prime }=\left( 11/40\right) V$. In particular, for $%
V=120,V^{\prime }=33$ one has $\alpha \simeq 1.68$ which is even larger than
the experimental value for (TMTSF)$_2$PF$_6$\cite{dar3}.

Consider now the case $V<2V^{\prime }$. It is shown from the Table that Eq.(%
\ref{vvv}) is violated in this case, at least, for large enough $V$ and $%
V^{\prime }$. As it was discussed above the reason is the formation of
bipolarons. There are two evidences of the coupling of electrons in the
pairs for large enough $2V^{\prime }-V$. First of all, the energy gap
defined as $\Delta =E\left( N+1\right) +E\left( N-1\right) -2E\left(
N\right) $ is finite (Fig. 1);

\begin{figure}[htbp]
   \includegraphics[bb =  103 189 532 617, angle=0, width=\textwidth]{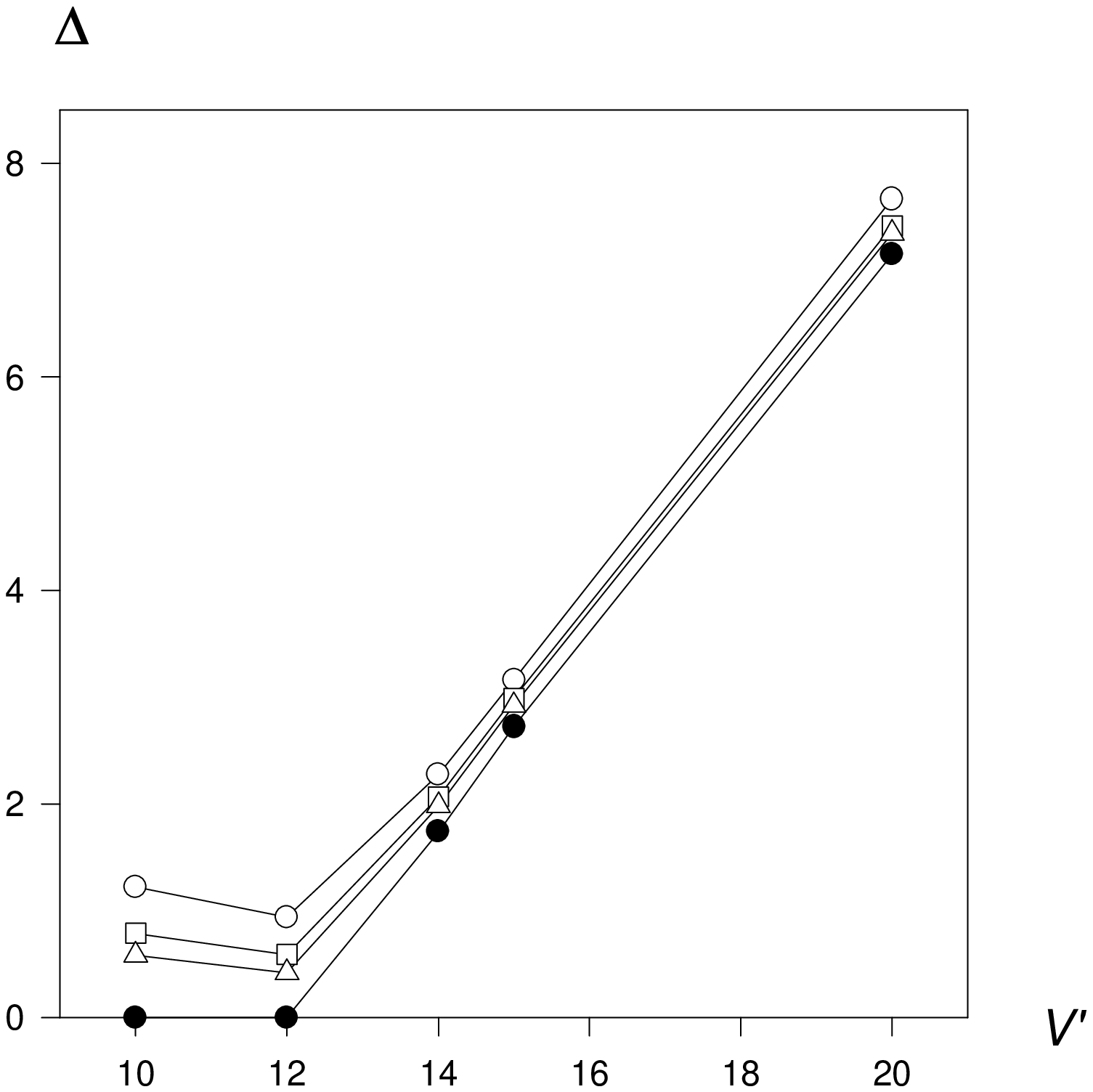}
\caption{Energy gap as a function of $V^{\prime }$ at $V=20$
for $L=$14($\circ$), 21({\small $\Box$}), 28({\footnotesize $\triangle$}),
and the results of the extrapolation to $L\to\infty$($\bullet$).}
\label{Fig1}
\end{figure}

\noindent at the same time, the Drude weight is also
non-zero which means that $\Delta $ is the superconducting, or bipolaron,
gap but not insulating one. Then, the ground state energy appears to be
oscillate as a function of magnetic flux in the ring $\Phi $ with the period
corresponding to the charge 2$e$ of current carriers. The corresponding
results are presented in Fig. 2,

\begin{figure}[htbp]
   \includegraphics[bb = 48 156 491 578, angle=0, width=\textwidth]{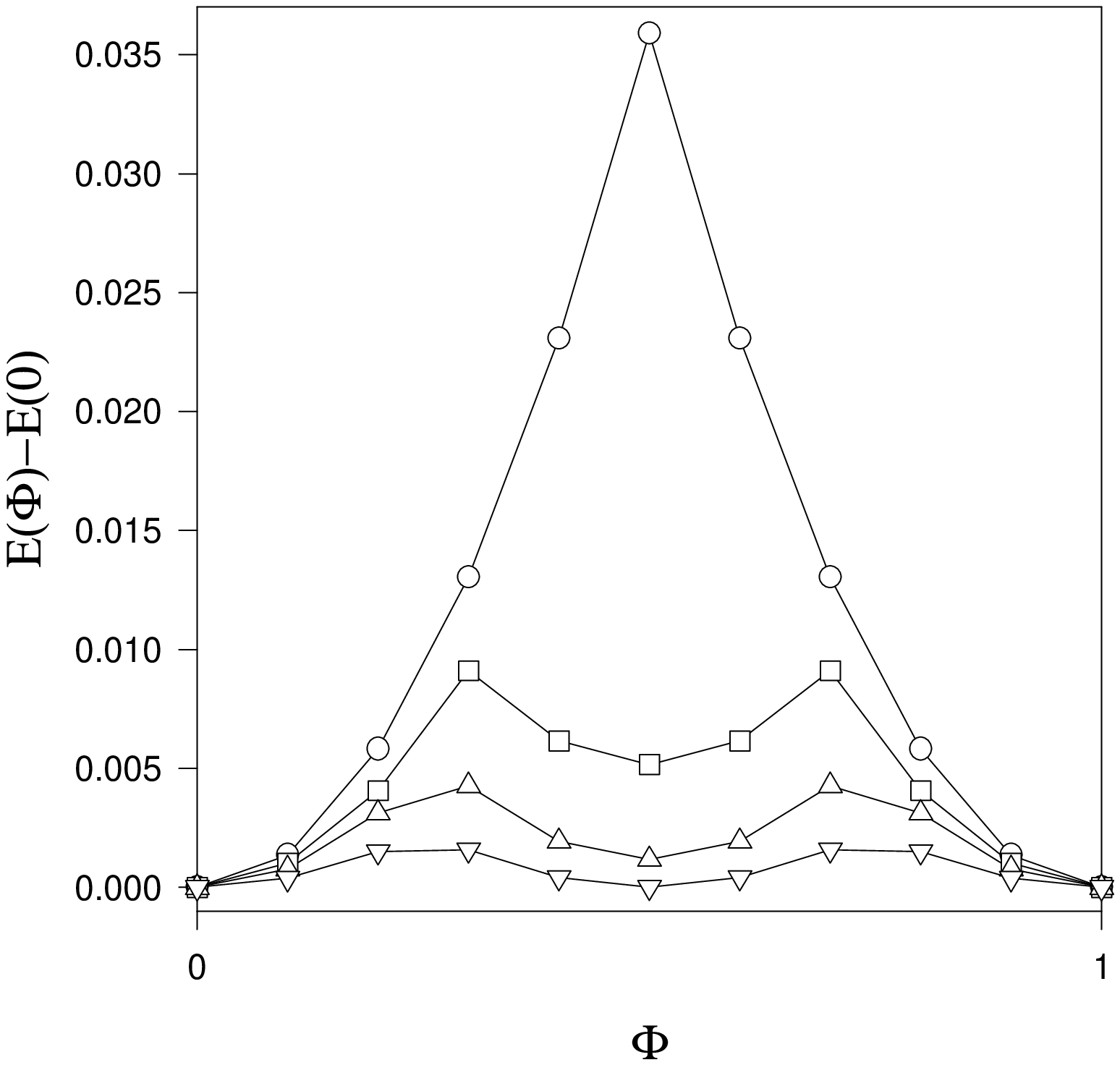}
\caption{Flux dependence of the total energy of the ring
$\left( L=21\right)$ for $V=20$ and different $V^{\prime }$: $V^{\prime}=12$
(diminished by a factor 2)($\circ$);
$V^{\prime }$=14({\small $\Box$}); $V^{\prime }$=15({\footnotesize $\triangle$});
$V^{\prime }$=20({\footnotesize $\bigtriangledown$}).}
\label{Fig2}
\end{figure}

the flux being introduced by the
replacement $t\rightarrow t\exp \left( 2\pi i\Phi \right) $ in Eq.(\ref{Hsp2}%
). According to these criteria, bipolarons do not form at small enough $%
2V^{\prime }-V\simeq t.$ To understand better this transition region we fit
the total energy by the expression
\begin{equation}
E\left( \Phi \right) -E\left( 0\right) =\beta \left( 1-\cos 2\pi \Phi
\right) +\gamma \left( 1-\cos 4\pi \Phi \right)   \label{alfabeta}
\end{equation}
where the first term corresponds to the ``normal'' current carriers and the
second one to the bipolarons. The fitted parameters $\beta$ and $\gamma$
are shown in Fig. 3.

\begin{figure}[htbp]
   \includegraphics[bb = 88 189 525 578, angle=0, width=\textwidth]{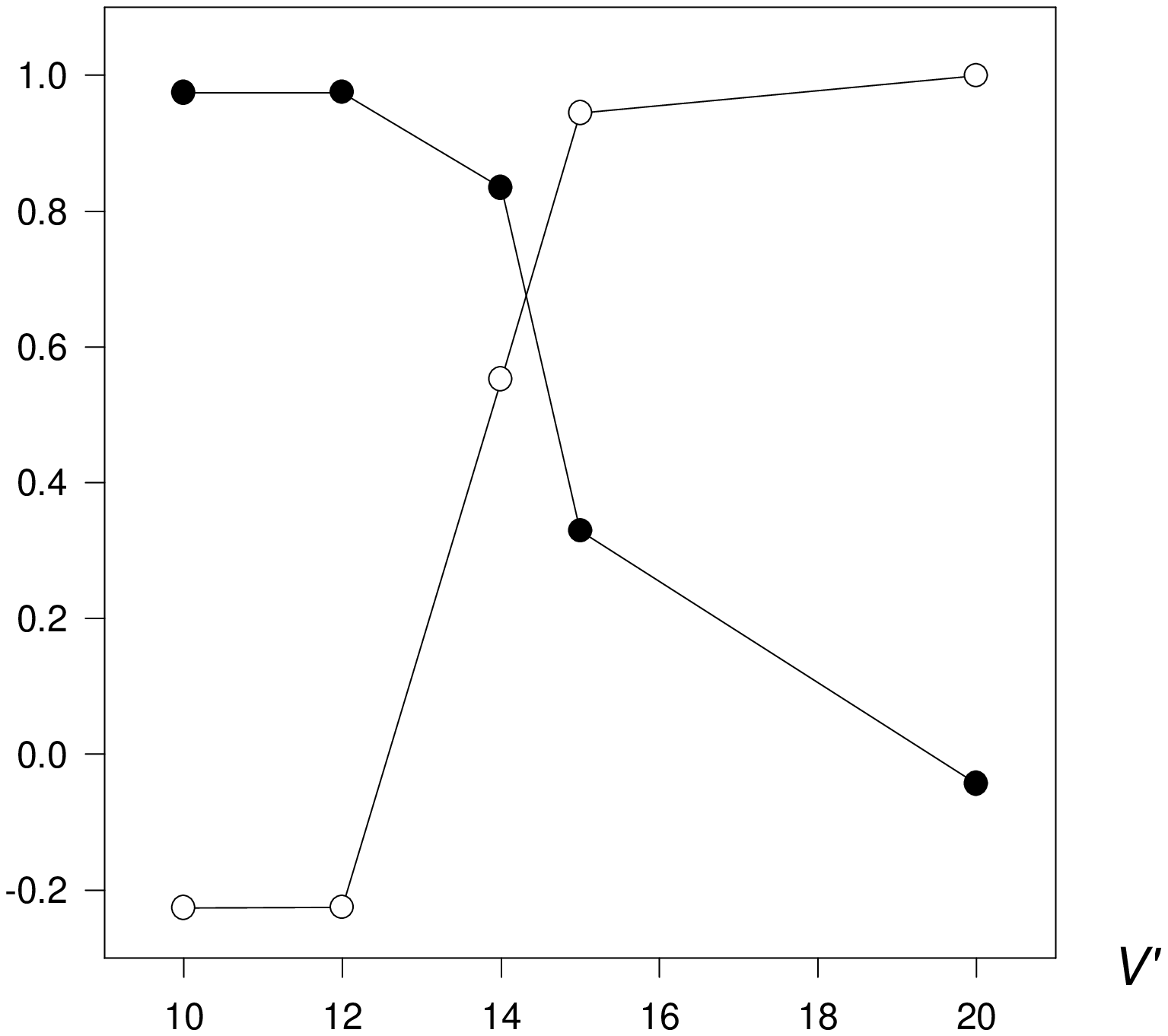}
\caption{Fitted parameters $\beta/\sqrt{\beta^2+\gamma^2}$
($\bullet$)
and $\gamma/\sqrt{\beta^2+\gamma^2}$ ($\circ$) from
Eq.(\ref{alfabeta}) as a function of $V^{\prime }$ for $V=20$ and $L=21.$}
\label{Fig3}
\end{figure}

One can see that there is a coexistence of bipolaron
conductivity and the usual one in the transition region $V\simeq 2V^{\prime
}\gg t$. It would be very interesting to investigate an explicit nature of
this coexistence (phase separation or a homogeneous state) but for this aim
computations for larger clusters are necessary.

In the limit of small $t$ one can simplify the problem using the operator
perturbation theory developed first by Bogoliubov\cite{bogol}. In the lowest
order in $t/\left( 2V^{\prime }-V\right) $ one can obtain
\begin{equation}
H_{eff}=-\frac{t^2}{2V^{\prime }-V}{\cal P}\sum_{i=1}^L\left( c_i^{\dagger
}c_{i+1}c_{i-1}^{\dagger }c_i+c_i^{\dagger }c_{i-1}c_{i+1}^{\dagger
}c_i\right) {\cal P}.  \label{Heff}
\end{equation}
where ${\cal P}$ is the projection operator on the state with the lowest
Coulomb energy. This Hamiltonian describes the bipolaron motion with the
effective transfer integral $t^2/\left( 2V^{\prime }-V\right) $. Our
numerical results give for the Drude weight
\begin{equation}
v_J=2\pi D\simeq \frac{2.3t^2}{2V^{\prime }-V}  \label{vJ}
\end{equation}
For small enough $t$ the latter can be arbitrarily small. This differ
drastically the case under consideration from the usual Verwey transition.
One can show for $\rho =1/2,V^{\prime }=0$ that near the transition point $%
V=2t$ the Drude weight has a finite jump\cite{SS}
\begin{equation}
v_J(V=2t-0)=\pi /2;\qquad v_J(V=2t+0)=0.  \label{jump}
\end{equation}

To conclude, we investigate numerically the ground state of 1D spinless
fermion model with competing interactions beyond half-filling. Depending on
the ratio of NN and NNN interactions it is either Luttinger liquid (with
unusually large exponent $\alpha $ which can be even larger than 1) or
bipolaron liquid. These results may be important to understand properties of
low-dimensional conductors; in particular, the value $\alpha >1$ has been
observed in photoemission experiments for (TMTSF)$_2$PF$_6$\cite{dar3}.

This work is partially supported by Russian Basic Research Foundation, grant
00-15-96544.


\end{document}